\documentstyle[12pt,epsf,amsfonts,amssymb,amsbsy]{article}
\newcommand\VV{\setbox0=\hbox{V}\hbox{\rm V\raise\ht0
  \hbox to0pt{\hss\vbox to0pt{\hbox{v}\vss}}}}
\def\slashchar#1{\setbox0=\hbox{$#1$}           
   \dimen0=\wd0                                 
   \setbox1=\hbox{/} \dimen1=\wd1               
   \ifdim\dimen0>\dimen1                        
      \rlap{\hbox to \dimen0{\hfil/\hfil}}      
      #1                                        
   \else                                        
      \rlap{\hbox to \dimen1{\hfil$#1$\hfil}}   
      /                                         
   \fi}                                         %

\begin{document}

\vspace*{4cm}

\begin{center}
{\large \bf Doubly heavy baryons in sum rules of NRQCD
}\\
\vspace*{5mm}
V.V. Kiselev$^{a)}$, A.I. Onishchenko$^{b)}$
\end{center}
\begin{center}
$^{a)}${\it State Research Center of Russia 
"Institute for High Energy Physics",\\ Protvino, Moscow region, 
142284 Russia}\\
\end{center}
\begin{center}
$^{b)}${\it Institute for Theoretical and Experimental Physics,\\ 
Moscow, 117218 Russia}.
\end{center}

\begin{abstract}
{The masses of baryons containing two heavy quarks and their couplings to the
corresponding quark currents are evaluated in the framework of NRQCD sum
rules. The coulomb-like corrections in the system of doubly heavy diquark are
taken into account, and the contribution of nonperturbative terms coming from
the quark, gluon and mixed condensates as well as the product of quark and
gluon condensates, is analyzed. The higher condensates destroy the
factorization of baryon and diquark correlators and provide the convergency of
sum rule method. As a result the accuracy of estimates is improved.}
\end{abstract}

\newpage

\section{Introduction}
Along with an experimental search for an explanation of electroweak symmetry
breaking and a physics beyond the Standard Model, high energies and
luminosities of particle accelerators running or being planned and under
construction provide a possibility to observe rare processes with heavy quarks.
A competitive topic here is a study of dynamics in flavored hadrons containing
two heavy quarks. It can play a fundamental role in an extraction of primary
parameters of quark interactions, since a distinction between the QCD effects
inside the doubly and singly heavy hadrons allows one strictly to constrain
incalculable nonperturbative quantities determining the isolation of pure
electroweak physics.

The real possibility of such experimental measurements was recently confirmed
by CDF Collaboration due to the first observation of $B_c$ meson \cite{cdf-bc}.
As predicted theoretically \cite{bc-rev}, this long-lived state of $\bar b$ and
$c$ quarks has the production cross sections, mass and decay rates, which
represent characteristic values for the doubly heavy hadrons. Thus, the
experimental search for the doubly heavy baryons can also be successful. Of
course, such the search would be more strongly motivated if it would be
supported by modern theoretical studies and evaluations of basic
characteristics for the doubly heavy baryons.

Some steps forward this program were already done. First, the production cross
sections of doubly heavy baryons in hadron collisions at high energies of
colliders and in fixed target experiments were calculated in the framework of
perturbative QCD for the hard processes and factorization of soft term related
to the nonperturbative binding the heavy quarks \cite{prod}. Second, the
lifetimes and branching fractions of basic decay modes were evaluated in the
Operator Product Expansion combined with the effective theory of heavy quarks,
which results in series over the inverse heavy quark masses and relative
velocities of heavy quarks inside the doubly heavy diquark \cite{ltime}.
Third, the families of doubly heavy baryons, which contain a set of narrow
excited levels in addition to the basic state, were described in the framework
of potential models \cite{pot}, so that the picture of spectra is very similar
to that of heavy quarkonia. Fourth, the QCD sum rules \cite{SVZ} were explored
for the two-point baryonic currents in order to calculate the masses and
couplings of doubly heavy baryons \cite{QCDsr}. However, the latter analysis
contains some disadvantages, which are related to instability of sum rules in a
region of parameters defining the baryonic currents. This fact leads to quite
large uncertainties in the calculations.

In the present paper we analyze the NRQCD sum rules for the two-point
correlators of currents corresponding to the doubly heavy baryons. The basic
physical motivation of such consideration is a nonrelativistic motion of two
heavy quarks inside a small size diquark bound with a light quark. This fact
leads to the definite expressions for the structure of baryonic currents
written down in terms of nonrelativistic heavy quarks. To the leading order of
inverse heavy quark mass and relative velocity of heavy quarks inside the
diquark, the NRQCD sum rules require the account for hard gluon corrections to
relate the nonrelativistic heavy quark correlators to the full QCD ones. The
corresponding anomalous dimensions of baryonic currents were calculated with
the two-loop accuracy \cite{anom}. The NRQCD structure of currents corresponds
to a definite choice of parameters in the full QCD expressions. Those values of
parameters are inside the instability region, observed in the analysis done
previously \cite{QCDsr}. We find the simple physical reason for this loose of
convergency: the behaviour of quantities versus the sum rule parameters (the
Borel variable or the moment number of spectral density) is determined by the
presence of doubly heavy diquark inside the baryon and, hence, the difference
between the masses of baryon and diquark. This mass difference takes a dominant
role unless we introduce the corrections corresponding to the essential
nonperturbative interactions between the doubly heavy diquark and the light
quark composing the baryon. In NRQCD sum rules this introduction is realized in
terms of nonperturbative condensates caused by higher dimension operators. We
show that the stability of sum rules can be reached due to the account for the
product of quark and gluon condensates in addition to the quark, gluon and
mixed condensates. This product was not taken into account in the previous
analysis in full QCD. Moreover, we carefully take into account the coulomb-like
$\alpha_s/v$ corrections inside the heavy diquark, which enhances the relative
weight of perturbative parts with respect to the condensate terms in the
calculated correlators.

In Section 2 we define the currents and represent the spectral densities in the
NRQCD sum rules for various operators included into the consideration. Section
3 is devoted to the numerical estimates. We find the masses of basic states,
which are close to the values obtained in the potential models. The results are
summarized in Conclusion.

\section{NRQCD sum rules for doubly heavy baryons}

\subsection{Baryonic Currents}

The currents of baryons with two heavy quarks $\Xi_{cc}^{\diamond}$,
$\Xi_{bb}^{\diamond}$ and $\Xi^{\prime \diamond}_{bc}$, where $\diamond$
denotes various electric charges depending on the flavour of light quark, are
associated with the spin-parity quantum numbers $j^P_d=1^+$ and $j^P_d=0^+$ for
the heavy diquark system with the symmetric and antisymmetric flavor structure,
respectively. Adding the light quark to the heavy quark system, one obtains
$j^P=\frac{1}{2}^+$ for the $\Xi^{\prime \diamond}_{bc}$ baryons and the pair
of degenerate states $j^P=\frac{1}{2}^+$ and $j^P=\frac{3}{2}^+$ for the
baryons $\Xi_{cc}^{\diamond}$, $\Xi_{bc}^{\diamond}$, $\Xi_{bb}^{\diamond}$ and
$\Xi_{cc}^{*\diamond}$, $\Xi_{bc}^{*\diamond}$, $\Xi_{bb}^{*\diamond}$. The
structure of baryonic currents with two heavy quarks is generally chosen as
\begin{equation}
J = [Q^{iT}C\Gamma\tau Q^j]\Gamma^{'}q^k\varepsilon_{ijk}.
\end{equation}
Here $T$ means transposition, $C$ is the charge conjugation matrix
with the properties $C\gamma_{\mu}^TC^{-1} = -\gamma_{\mu}$ and
$C\gamma_5^TC^{-1} = \gamma_5$, $i,j,k$ are colour indices and $\tau$ is a
matrix in the flavor space. The effective static field of the heavy quark is
denoted by $Q$. To the leading order over both the relative velocity of heavy
quarks and their inverse masses, this field contains the ``large'' component
only in the hadron rest frame.

Here, unlike the case of baryons with a single heavy quark \cite{yakov}, there
is the only independent current component $J$ for each of the ground state
baryon currents. They equal
\begin{eqnarray}
J_{\Xi^{\prime \diamond}_{QQ^{\prime}}} &=& [Q^{iT}C\tau\gamma_5
Q^{j\prime}]q^k\varepsilon_{ijk},\nonumber\\
J_{\Xi_{QQ}^{\diamond}} &=& [Q^{iT}C\tau\boldsymbol{\gamma}^m
Q^j]\cdot\boldsymbol{\gamma}_m\gamma_5
q^k\varepsilon_{ijk},
\label{def}\\
J_{\Xi_{QQ}^{*\diamond}}^n &=& [Q^{iT}C\tau\boldsymbol{\gamma}^n
Q^j]q^k\varepsilon_{ijk}+\frac{1}{3}\boldsymbol{\gamma}^n
[Q^{iT}C\boldsymbol{\gamma}^m
Q^j]\cdot\boldsymbol{\gamma}_m q^k\varepsilon_{ijk},\nonumber
\end{eqnarray}
where $J_{\Xi_{QQ}^{*\diamond}}^n$ satisfies the spin-3/2 condition
$\boldsymbol{\gamma}_n J_{\Xi_{QQ}^{*\diamond}}^n = 0$. The flavor matrix
$\tau$ is antisymmetric for $\Xi^{\prime \diamond}_{bc}$ and symmetric for
$\Xi_{QQ}^{\diamond}$ and $\Xi_{QQ}^{*\diamond}$. The currents written down in
Eq. (\ref{def}) are taken in the rest frame of hadrons. The corresponding
expressions in a general frame moving with a velocity four-vector $v^{\mu}$ can
be obtained by the substitution of $\boldsymbol{\gamma}^m\to
\gamma_{\perp}^{\mu}=\gamma^{\mu}-\slashchar{v} v^{\mu}$.

To compare with the full QCD analysis we represent the expression for the
$J_{\Xi^{\prime \diamond}_{bc}}$ current given in \cite{QCDsr}
$$
J_{\Xi^{\prime \diamond}_{bc}} = \{r_1 [u^{iT}C\gamma_5 
c^{j}]b^k + r_2 [u^{iT}C c^{j}]\gamma_5 b^k + r_3 [u^{iT}C\gamma_5
\gamma_mu c^{j}]\gamma^\mu b^k\}\varepsilon_{ijk},
$$
so that the NRQCD structure can be obtained by the choice of $r_1=r_2=1$ and
$r_3=0$ and the antisymmetric permutation of $c$ and $b$ flavors. As we have
already mentioned in the Introduction, the authors of \cite{QCDsr} reported
that the convergency of OPE is ``bad'' in the region around the point given by
the NRQCD parameters. The instability results in huge uncertainties of
estimates. To get rid this disadvantage, we analyze the NRQCD sum rules in
details.

\subsection{Description of the method}

In this section we describe steps required for the evaluation of two-point
correlation functions in the NRQCD approximation and the connection to physical
characteristics of doubly heavy baryons. We start from the correlator of two
baryonic currents with the half spin
\begin{equation}
\Pi(t) = i\int d^4x e^{ipx}\langle 0|T\{J(x),\bar
J(0)\}|0\rangle = \slashchar{v} F_1(t) + F_2(t),\label{2pcor}
\end{equation}
where $t = k\cdot v$, $k_{\mu}$ and $p_{\mu}$ denote the residual and full
momenta of doubly heavy baryon, respectively, $v_{\mu}$ is its four-velocity,
which are related by the following formula:
\begin{equation}
p_{\mu} = k_{\mu} + (m_1+m_2)v_{\mu},
\end{equation}
where $m_1$ and $m_2$ are the heavy quark masses. The appropriate definitions
of scalar formfactors for the 3/2-spin baryon are given by the following:
\begin{equation}
\Pi_{\mu\nu}(t) = i\int d^4x e^{ipx}\langle 0|T\{J_\mu (x),\bar
J_\nu (0)\}|0\rangle = -g_{\mu\nu}[\slashchar{v} \tilde F_1(t) +
\tilde F_2(t)]+\ldots,\label{2vcor}
\end{equation}
where we will not concern for contributions with distinct lorentz structures.
The scalar correlators $F$ can be evaluated in deep euclidean region by
employing the Operator Product Expansion (OPE) in the framework of NRQCD for
the time ordered product of currents in Eqs.(\ref{2pcor}), (\ref{2vcor}), say,
\begin{equation}
F_{1,2}(t) = \sum_d C^{(1,2)}_d(t)O_d,
\end{equation}
where $O_d$ denotes the local operator with a given dimension $d$:
$O_0 = \hat 1$, $O_3 = \langle\bar qq\rangle$, $O_4 =
\langle\frac{\alpha_s}{\pi}G^2\rangle$, \ldots,
and the functions $C_d(t)$ are the corresponding Wilson coefficients of OPE.
In this work we include the nonperturbative contributions given by quark, gluon
and mixed condensates. So, evaluating the contribution of quark condensate
operator, we use the following OPE for the correlator of two quark fields
\cite{Smilga}:
\begin{eqnarray}
\langle 0|T\{q_i^a(x)\bar q_j^b(0)|0\rangle &=& -\frac{1}{12}
\delta^{ab}\delta_{ij}\langle\bar q  q \rangle\cdot [1 + \frac{m_0^2x^2}{16}
+ \frac{\pi^2x^4}{288}\langle\frac{\alpha_s}{\pi}G^2\rangle + ... ],
\label{qq}
\end{eqnarray}
where the value of mixed condensate is parameterized by introducing the
variable $m_0^2$, which is numerically determined as $m_0^2 \approx 0.8$
GeV$^2$.

For the sake of simplicity, we write down the Wilson coefficients of unity and
quark-gluon operators by making use of dispersion relations over $t$,
\begin{equation}
C_d(t) = \frac{1}{\pi}\int_0^{\infty}\frac{\rho_d(\omega)d\omega}{\omega - t},
\end{equation}
where $\rho$ denotes the imaginary part in the physical region
of NRQCD. Thus, the calculation of Wilson coefficients of operators under
consideration is transformed into the evaluating the corresponding spectral
densities.

To relate the NRQCD correlators to the real hadrons, we use the dispersion 
representation for the two point function with the physical spectral
density, given by the appropriate resonance and continuum part. The coupling
constants of baryons are defined by the following expressions:
\begin{eqnarray}
\langle 0| J(x)|{\Xi_{QQ}^{\diamond}}(p)\rangle & = & i
Z_{\Xi_{QQ}^{\diamond}} u(v,M_{\Xi}) e^{ip x}, \nonumber \\
\langle 0| J^m(x) |{\Xi_{QQ}^{*\diamond}}(p,\lambda)\rangle & = & i
Z_{\Xi_{QQ}^{*\diamond}} u^m(v,M_{\Xi}) e^{ip x}, \nonumber 
\end{eqnarray}
where the spinor field with four-velocity $v$ and mass $M_\Xi$ satisfies the
equation $\slashchar{v} u(v,M_{\Xi}) = u(v,M_{\Xi})$, and 
$u^m(v,M_{\Xi})$ denotes the transversal spinor, so that $(\gamma^m-v^m
\slashchar{v}) u^m(v,M_{\Xi}) =0$.

We suppose that the continuum density, starting from the threshold
$\omega_{cont}$ is equal to that of calculated in the framework of NRQCD. Then
in sum rules equalizing the correlators, calculated in NRQCD and given by the
physical states, the integrations above $\omega_{cont}$ cancel each other
in two sides of relation. This fact leads to the dependence of calculated
masses and couplings on the value of $\omega_{cond}$. Further, we write down
the correlators at the deep under-threshold point of $t_0=-(m_1+m_2)+t$ with
$t\to 0$, which corresponds to the limit of $p^2\to 0$. The approximation of
single bound state leads to the following expression for the resonance
contribution:
\begin{equation}
F_{1,2}(t)= \frac{M_{\Xi}
|Z_{\Xi}|^2}{M^2_{\Xi}-t^2},
\end{equation}
which can be expanded in series \footnote{To the nonrelativistic approximation
we have to substitute for $\frac{1}{M_{\Xi}^2-t^2}$ by the single pole in the
physical region over $t$,  $\frac{1}{(M_\Xi-t)(M_\Xi+t)}\approx
\frac{1}{M_\Xi(M_\Xi-t)}$.} over $t$. Then, the sum rules state the following
equalities for the terms standing in
front of powers of $t$
\begin{equation}
\frac{1}{\pi}\int_0^{\omega_{cont}}\rho_{1,2} (\omega )d\omega
\frac{1}{(\omega + m_1 + m_2)^{n}} = |Z_{\Xi}|^2 \frac{1}{M_{\Xi}^{n}},
\end{equation}
where $\rho_j$ contains the contributions given by various operators in OPE for
the corresponding scalar correlators $F_j$.
Introducing the following notation for $n$-th moment of two point 
correlation function
\begin{equation}
{\cal M}_n	=\frac{1}{\pi}\int_0^{\omega_{cont}}\frac{\rho (\omega )d\omega}
{(\omega + m_1 + m_2)^{n+1}},
\end{equation}
we can obtain the estimates of baryon mass $M_{\Xi}$, for example, as the
following:
\begin{equation}
M_{\Xi}[n] = \frac{{\cal M}_n}{{\cal M}_{n + 1}},
\end{equation}
and the coupling is determined by the expression
\begin{equation}
|Z_{\Xi}[n]|^2 = {\cal M}_n M_{\Xi}^{n + 1},
\end{equation}
where we see the dependence of sum rule results on the scheme parameter.
Therefore, we tend to find the region of parameter values, where, first, the
result is stable under the variation of $n$, and, second, the both correlators
$F_1$ and $F_2$ reproduce equal values of physical quantities: masses and
couplings. The problem of full QCD was the second point: the difference between
the evaluated variables.

The similar procedure can be described in the Borel scheme, wherein the
consideration of ${\cal M}_n$ is replaced by the analysis of function ${\cal
B}[w]$, which is defined by 
\begin{equation}
{\cal B}[w]=\frac{1}{\pi}\int_0^{\omega_{cont}}{\rho (\omega )d\omega}
e^{-(\omega + m_1 + m_2)/w},
\end{equation}
equal to the resonance term given by
\begin{equation}
{\cal B}_{\Xi}[w]= |Z_{\Xi}|^2 e^{-M_{\Xi}/w}.
\end{equation}
Then, again the scheme dependence of masses and couplings appears because of
variation versus the Borel parameter $w$.

\subsection{Calculating the spectral densities}
\noindent
In this subsection we present analytical expressions for the perturbative
spectral functions in the NRQCD approximation. The evaluation of
spectral densities involves the standard use of Cutkosky rules 
\cite{Cutk} with some modifications motivated by NRQCD. We explore the
prescription that the discontinuity of two-point functions under
consideration can be evaluated using the following substitutions for heavy and
light quark propagators, correspondingly:
\begin{eqnarray}
{\rm heavy\; quark:}\;\;&& \frac{1}{p_0-(m+\frac{{\vec p}^2}{2m})}\to 2\pi
i\cdot\delta (p_0-(m+\frac{{\vec p}^2}{2m})), \nonumber\\
{\rm light\; quark:}\;\;&& \frac{1}{p^2-m^2} \to 2\pi i\cdot\delta
(p^2-m^2).\nonumber
\end{eqnarray}
For the perturbative spectral densities $\rho_{1,H} (\omega)$ and $\rho_{2,H}
(\omega)$ standing in front of unity operator in $F_1$ and $F_2$, respectively,
where $H = \Xi^{\prime \diamond}_{QQ^{\prime}}, \Xi_{QQ}^{\diamond}$ or
$\Xi_{QQ}^{*\diamond}$, we have the following expressions:
\begin{eqnarray}
\rho_{1,\Xi^{\prime\diamond}_{QQ^{\prime}}}(\omega) &=&
\frac{1}{15015\pi^3(\omega+m_1+m_2)^3}16\sqrt{2}\left (\frac{m_1m_2}{m_1+m_2}
\right )^{3/2}\omega^{7/2}[429m_1^3+429m_2^3+\nonumber\\
&&715m_2^2\omega+403m_2\omega^2+77\omega^3+143m_1^2(9m_2+5\omega)+
13m_1(99m_2^2+ \\
&& 110m_2\omega+33\omega^2)],\nonumber\\
\rho_{1,\Xi_{QQ}^{\diamond}}(\omega) &=& 3
\rho_{1,\Xi^{\prime\diamond}_{QQ^{\prime}}}(\omega) = 3
\rho_{1,\Xi_{QQ}^{*\diamond}}(\omega), \label{ss}\\
\rho_{2,\Xi_{QQ}^{\diamond}}(\omega) &=&
\rho_{2,\Xi^{\prime\diamond}_{QQ^{\prime}}}(\omega) =
\rho_{2,\Xi_{QQ}^{*\diamond}}(\omega)=0.
\end{eqnarray}
Note that in the leading order of theory with the effective heavy quarks their
spins are decoupled from the interaction, that causes the spin symmetry
relations given in Eq.(\ref{ss}). The factor 3 stands because of the
normalization of vector diquark current. We see also that in the leading
approximation of perturbative NRQCD the $F_2$ correlators are equal to zero.
This is because the interaction of massless light quark with the doubly heavy
diquark is switched off in this order, and there is no mass term structure in
the correlator.

The coulomb-like interaction inside the heavy diquark can be taken into account
by the introduction of Sommerfeld factor $\bf C$ for the diquark spectral
densities before the integration over the diquark invariant mass to obtain the
baryon spectral densities, so that
\begin{equation}
\rho_{diquark} = \rho_{diquark}^{B}\cdot{\bf C},
\end{equation}
with
\begin{equation}
{\bf C} = \frac{2\pi\alpha_s}{3v_{12}}\left [ 1- \exp\left
(-\frac{2\pi\alpha_s}{3v_{12}} \right)\right ]^{-1},
\end{equation}
where $v_{12}$ denotes the relative velocity of heavy quarks inside the
diquark, and we have taken into account the color anti-triplet structure of
diquark. The relative velocity is given by the following expression:
\begin{equation}
v_{12} = \sqrt{1-\frac{4m_1m_2}{Q^2-(m_1-m_2)^2}},
\end{equation}
where $Q^2$ is the square of heavy diquark four-momentum. In NRQCD we take the
limit of low velocities, so that denoting the diquark invariant mass squared by
$Q^2=(m_1+m_2+\epsilon)^2$ and the reduced quark pair mass by $m_{12}=m_1
m_2/(m_1+m_2)$, we find
$$
{\bf C} = \frac{2\pi\alpha_s}{3v_{12}},\;\;\;
v_{12}^2 = \frac{\epsilon}{2m_{12}},
$$
at $\epsilon\ll m_{12}$. The corrected spectral densities are equal to
\begin{eqnarray}
\rho^C_{1,\Xi^{\prime\diamond}_{QQ^{\prime}}}(\omega) &=&
\frac{\alpha_s}{16\pi^2 (\omega+m_1+m_2)^3}\left (\frac{m_1m_2}{m_1+m_2}
\right )^{2}\omega^{3}[2 m_1+2 m_2^3+\omega]^3,\nonumber\\
\rho^C_{1,\Xi_{QQ}^{\diamond}}(\omega) &=& 3
\rho^C_{1,\Xi^{\prime\diamond}_{QQ^{\prime}}}(\omega) = 3
\rho^C_{1,\Xi_{QQ}^{*\diamond}}(\omega). \label{css}
\end{eqnarray}

Further, the spectral functions, connected to the condensates of light quarks 
and gluons, can be derived. For the quark condensate term we have the following 
expressions:
\begin{eqnarray}
\rho_{2,\Xi^{\prime\diamond}_{QQ^{\prime}}}^{\bar qq}(\omega) &=&
-\frac{\sqrt{2}}{\pi}\left (\frac{m_1m_2}{m_1+m_2}\right)^{3/2}\sqrt{\omega},\\
\rho_{2,\Xi_{QQ}^{\diamond}}^{\bar qq}(\omega) &=& 3
\rho_{2,\Xi^{\prime\diamond}_{QQ^{\prime}}}^{\bar qq}(\omega) = 3
\rho_{2,\Xi_{QQ}^{*\diamond}}^{\bar qq}(\omega), \\
\rho_{1,\Xi_{QQ}^{\diamond}}^{\bar qq}(\omega) &=& 
\rho_{1,\Xi^{\prime\diamond}_{QQ^{\prime}}}^{\bar qq}(\omega) = 
\rho_{1,\Xi_{QQ}^{*\diamond}}^{\bar qq}(\omega)=0,
\end{eqnarray}
which have to be multiplied by the Sommerfeld factor $\bf C$, wherein the
variable $\epsilon$ is substituted by $\omega$, since in this case we have no
integration over the quark-diquark invariant mass.

It is interesting to stress that in NRQCD the light quark condensate
contributes to the $F_2$ correlators, only. This fact has a simple physical
explanation: to the leading order the light quark operator can be factorized in
the expression for the correlator of baryonic currents. Indeed, we can write
down for the condensate contribution
$$
\langle 0|T\{J(x),\bar J(0)\}|0\rangle = \langle 0|T\{q_i^a(x)\bar
q_i^a(0)|0\rangle\cdot \frac{\hat 1}{12}\cdot \langle 0|T\{J^j_d(x),\bar
J^j_d(0)\}|0\rangle +\ldots,
$$
where $J^j_d(x)$ denotes the appropriate diquark current with the color index
$j$, as it is defined by the baryon structure in Eqs. (\ref{def}). So, we see
that the restriction by the first term independent of $x$ in the expansion for
the quark correlator in (\ref{qq}) results in the independent contribution of
diquark correlator to the baryonic one. Then, since the diquark correlator is
isolated in $F_2$ from the baryonic formfactor $F_1$, the NRQCD sum rules lead
to the evaluation of diquark mass and couplings from $F_2$, and estimation of
baryon masses and couplings from $F_1$. These masses and couplings are
different. The positive point is the possibility to calculate the binding
energy for the doubly heavy baryons $\bar \Lambda = M_\Xi - M_{diquark}$. The
disadvantage is the instability of NRQCD sum rules at this stage, since the
various formfactors or correlators lead to the different results. In  sum rules
of full QCD various choices of parameters in the definitions of baryonic
currents result in an admixture of pure diquark correlator in various
formfactors, so that the estimations acquire huge uncertainties. Say, the
characteristic ambiguity in the evaluation of baryon mass in full QCD is about
300 MeV, i.e. the value close to the expected estimate of $\bar \Lambda$. The
analysis in the framework of NRQCD makes this result to be not unexpectable.
Moreover, it is quite evident that the introduction of interactions between the
light quark and the doubly heavy diquark destroys the factorization of diquark
correlator. Indeed, we see that due to the higher terms in expansion
(\ref{qq}), the diquark factorization is explicitly broken, which has to result
in the convergency of estimates obtained from $F_1$ and $F_2$. Below we show
numerically that this fact is valid. Technically, we point out that the
contribution to the moments of spectral density, determined by the light quark
condensate including the mixed condensate and the product of quark and gluon
condensates, can be calculated after the exploration of (\ref{qq}), so that
\begin{equation}
{\cal M}_n^{q\bar q} = \left[1-\frac{(n+2)!}{n!}\frac{m_0^2}{16} +
\frac{(n+4)!}{n!} \frac{\pi^2}{288} \langle\frac{\alpha_s}{\pi}G^2\rangle
\right]\; {\cal M}_n^{\langle\bar q q\rangle}.
\end{equation}
For the corrections due to the gluon condensate we have
\begin{eqnarray}
\rho_{1,\Xi^{\prime\diamond}_{QQ^{\prime}}}^{G^2}(\omega) &=&
\frac{\sqrt{\frac{m_1m_2}{m_1 + m_2}\omega}}{1344\sqrt{2}\pi(m_1 + m_2)^2
(\omega + m_1 + m_2)^3}\cdot (28m_1^2+41m_1m_2+28m_2^2)\cdot \nonumber\\
&& (\omega^3-7m_1^3+7m_1^2\omega-21m_1^2m_2+7m_2\omega^2+7m_2^2\omega
-7m_2^3+7m_1\omega^2\\
&& +14m_1m_2\omega-21m_1m_2^2),\nonumber\\
\rho_{1,\Xi_{QQ}^{\diamond}}^{G^2}(\omega) &=& 3
\rho_{1,\Xi^{\prime\diamond}_{QQ^{\prime}}}^{G^2}(\omega) = 3
\rho_{1,\Xi_{QQ}^{*\diamond}}^{\bar qq}(\omega), \\ 
\rho_{2,\Xi_{QQ}^{\diamond}}^{G^2}(\omega) &=& 
\rho_{2,\Xi^{\prime\diamond}_{QQ^{\prime}}}^{G^2}(\omega) = 
\rho_{2,\Xi_{QQ}^{*\diamond}}^{\bar qq}(\omega)=0, 
\end{eqnarray}
which are written down for $O_4$, having the form $O_4 =
\langle\frac{\alpha_s}{\pi}G^2\rangle$.

For the product of condensates $\langle\bar qq\rangle\langle
\frac{\alpha_s}{\pi}G^2\rangle$, wherein the gluon fields are connected to the
heavy quarks in contrast to the light quark, we have computed the contribution
to the two-point correlation function itself. It has the following form:
\begin{eqnarray}
F_{2,\Xi^{\prime\diamond}_{QQ^{\prime}}}^{\bar qqG^2}(\omega) &=& 
-\frac{\pi\sqrt{\frac{m_1m_2}{m_1 + m_2}}(7m_1^2 + 8m_1m_2 + 7m_2^2)}
{3072\sqrt{2}(-\omega)^{5/2}},\\
F_{2,\Xi_{QQ}^{\diamond}}^{\bar qqG^2}(\omega) &=& 3
F_{2,\Xi^{\prime\diamond}_{QQ^{\prime}}}^{\bar qqG^2}(\omega) = 3
F_{2,\Xi_{QQ}^{*\diamond}}^{\bar qqG^2}(\omega),\\ 
F_{1,\Xi_{QQ}^{\diamond}}^{\bar qqG^2}(\omega) &=& 
F_{1,\Xi^{\prime\diamond}_{QQ^{\prime}}}^{\bar qqG^2}(\omega) = 
F_{1,\Xi_{QQ}^{*\diamond}}^{\bar qqG^2}(\omega)=0,
\end{eqnarray}
where $\omega=-(m_1+m_2)+t$ at the point under consideration, and the
correlators have to be expanded in series of t, which gives the moments.

Thus, we provide the NRQCD sum rules, where we take into account the
perturbative terms and the vacuum expectations of quark-gluon operators up to
the contributions by the light quark condensate, gluon condensate, their
product and the mixed condensate. Note, that the product of condensates is
essential for the doubly heavy baryons, and we present the full NRQCD
expression for this term, including the interaction of nonperturbative gluons
with both the light and heavy quarks. The correct introduction of coulomb-like
interactions is done for the perturbative spectral densities of heavy diquark,
which is important for the nonrelativistic heavy quarks. Finally, we find the
spin-symmetry relation for the baryon couplings in NRQCD
$$
|Z_\Xi|^2 = 3 |Z_{\Xi\prime}|^2 = 3 |Z_{\Xi^*}|^2.
$$

\subsection{Anomalous dimensions for the baryonic currents}

To connect the NRQCD sum rules to the quantities in full QCD we have to take
into account the anomalous dimensions of effective baryonic currents with the
nonrelativistic quarks. They determine the factors, which have to multiply the
NRQCD correlators to obtain the values in full QCD. Indeed, to the leading
order of NRQCD we have the relation
$$
J^{QCD} = C_J(\alpha_s,\mu) \cdot J^{NRQCD},
$$
where the coefficient $C_J$ depends on the normalization scale $\mu$ and obeys
the matching condition at the starting point of $\mu_0 = m_1+m_2$. The
anomalous dimensions of NRQCD currents are independent of the diquark spin
structure in the leading order. They are equal to \cite{anom}
\begin{eqnarray}
\gamma &=& \frac{d\ln C_J(\alpha_s,\mu)}{d\ln (\mu)} =
\sum_{m=1}^\infty\left(\frac{\alpha_s}{4\pi}\right)^m\gamma^{(m)},
	   \nonumber\\
\gamma^{(1)} &=& \Big(-2C_B(3a-3)+3C_F(a-2)\Big), \label{gamma}\\
\gamma^{(2)} &=& \frac{1}{6}(-48(-2+6\zeta (2))C_B^2+C_A((104-240\zeta
(2))C_B-101C_F) \nonumber \\
&& -64C_BN_FT_F+C_F(-9C_F+52N_FT_F)),\nonumber
\end{eqnarray}
where $C_F=(N_c^2-1)/2N_c$, $C_A=N_c$, $C_B=(N_c+1)/2N_c$, and $T_F=1/2$ for 
$N_c=3$, $N_F$ being the number of light quarks. In Eq.(\ref{gamma}) we give
the one-loop result with the arbitrary gauge parameter $a$, and the two-loop
anomalous dimension is represented in the Feynman gauge $a=1$. 
So, numerically at $N_F=3$ and $a=1$ we find
\begin{equation}
\gamma^{(1)} = -4,\;\;\;
\gamma^{(2)} \approx -188.24.
\end{equation}
In the leading logarithmic approximation and to the one-loop accuracy, the
coefficient $C_J$ is given by the
expression
\begin{equation}
C_J(\mu) = \left(
\frac{\alpha_s(\mu_0)}{\alpha_s(\mu)}\right)^{\frac{\gamma^{(1)}}{2\beta_0}}
\end{equation}
where $\beta_0 = 11N_c/3 - 2 N_F/3 =9$. To evaluate the two-loop expression for
$C_J$ we have to know subleading corrections in the first $\alpha_s$ order in
addition to the anomalous dimensions. These corrections are not available yet,
so we restrict ourselves by the one-loop accuracy.

Further, we have to determine the normalization point for the NRQCD estimates
$\mu=\mu_1$. We put it to the average momentum transfer inside the doubly heavy
diquark, so that $\mu_1^2=T_d 2 m_{12}$, where $T_d$ denotes the kinetic energy
in the system of two heavy quarks, which is phenomenologically independent of
the quark flavors and approximately equal to 0.2 GeV. Then, the coefficients
$C_J$ are equal to
\begin{equation}
C_{\Xi_{cc}} \approx 1.95,\;\;\;
C_{\Xi_{bc}} \approx 1.52,\;\;\;
C_{\Xi_{bb}} \approx 1.30,
\end{equation}
with the characteristic uncertainty about 10\% because of the variation of
initial and final points $\mu_{0,1}$.

Finally, we emphasize that the values of $C_J$ do not change the estimates of
baryon masses calculated in the sum rules of NRQCD. However, they are essential
in the evaluation of baryon couplings, which acquire these multiplicative
factors.

\section{Numerical estimates}

In the present paper we explore the sum rule scheme of moments for the spectral
functions of correlators. We have to stress that in this scheme the dominant
uncertainty in the results is caused by the variation of heavy quark masses. In
the analysis we chose the following region of mass values:
\begin{equation}
m_b = 4.6-4.7\; {\rm GeV,}\;\;\;\; 
m_c = 1.35-1.40\; {\rm GeV,}
\end{equation}
which is ordinary used in the sum rule estimates for the heavy quarkonia. Next
critical point is the value of QCD coupling constant determining the
coulomb-like interactions inside the doubly heavy diquark. Indeed, it stands
linearly in front of the perturbative functions of diquark contributions. Thus,
the introduction of $\alpha_s/v$-corrections is essential for both the baryon
couplings and the relative contributions of perturbative terms and condensates
to the baryon masses. To decrease the uncertainty we impose the same approach
to the heavy quarkonia, where it is well justified, and then, we extract the
characteristic values for the heavy-heavy systems from the comparison of
calculations with the current data on the leptonic constants of heavy
quarkonia, which are known experimentally for $c\bar c$ and $b\bar b$ or
evaluated in various approaches for $\bar b c$. So, our calculations give the
following couplings of coulomb interactions
\begin{equation}
\alpha_s(b\bar b) = 0.37,\;\;\;
\alpha_s(c\bar b) = 0.45,\;\;\;
\alpha_s(c\bar c) = 0.60.\;\;\;
\end{equation}
The dependence of estimates on the value of thershold for the continuum
contribution is not so valuable as on the quark masses. We fix the region of
$\omega_{cont}$ as
\begin{equation}
\omega_{cont} = 1.3-1.4\; {\rm GeV.}
\end{equation}
For the condensates of quarks and gluons the following regions are under
consideration:
\begin{equation}
\langle\bar q  q \rangle = -(250-270\;{\rm MeV})^3,\;\;\;
m_0^2 = 0.75- 0.85\;{\rm GeV}^2,\;\;\;
\langle\frac{\alpha_s}{\pi}G^2\rangle = (1.5-2)\cdot 10^{-2}\;{\rm GeV}^4.
\end{equation}
So, we have described the choices of parameters.

Figs. \ref{mcc}-\ref{mbb} represent the calculated difference of masses
extracted from the $F_1$ and $F_2$ correlators\footnote{In these figures we
have fixed the value of gluon condensate $\langle\frac{\alpha_s}{\pi}G^2\rangle
= 1.7\cdot 10^{-2}\;{\rm GeV}^4$ and arranged $m_0^2$ in the above region to
reach zero differences between the masses, though the variation of parameters
leads to errors in the estimates quoted below.} for the baryons $\Xi_{cc}$, 
$\Xi_{bc}$ and $\Xi_{bb}$. We certainly see that at low numbers of moments for
the spectral densities, the baryon-diquark mass difference can be evaluated as
\begin{equation}
\bar \Lambda = 0.40\pm 0.03\; {\rm GeV,}
\end{equation}
which is quite a reasonable value, being in a good agreement with the estimates
in the heavy-light mesons. In the region of mass difference stability we can
fix the number of moment for the spectral density, say, $n=27\pm 1$ for
$\Xi_{bc}$, and calculate the corresponding masses of baryons, which are equal
to 
\begin{equation}
M_{\Xi_{cc}} = 3.47\pm 0.05\; {\rm GeV},\;\;\;
M_{\Xi_{bc}} = 6.80\pm 0.05\; {\rm GeV},\;\;\;
M_{\Xi_{bb}} = 10.07\pm 0.09\; {\rm GeV},\;\;\;
\end{equation}
where we do not take into account the spin-dependent splitting caused by the
$\alpha_s$-corrections to the heavy-light interactions, which are not available
yet. The uncertainties in the mass values are basically given by the variation
of heavy quark masses. The convergency of NRQCD sum rules allows one to improve
the accuracy of estimates in comparison with the previous analysis in full QCD
\cite{QCDsr}. The obtained values are in agreement with the calculations in the
framework of nonrelativistic potential models \cite{pot}.

\setlength{\unitlength}{1mm}
\begin{figure}[th]
\begin{center}
\begin{picture}(100,80)
\put(0,0){\epsfxsize=10cm \epsfbox{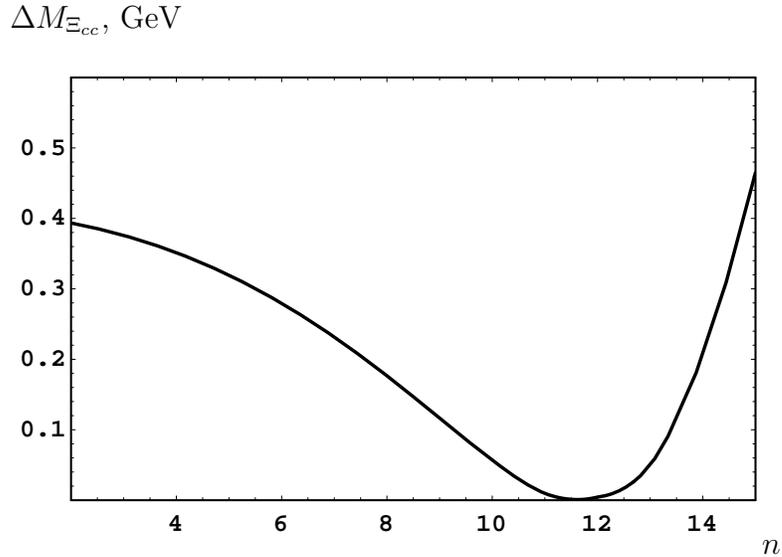}}
\put(100,0){$n$}
\put(0,70){$\Delta M_{\Xi_{cc}}$, GeV}
\end{picture}
\end{center}
\caption{The difference between the $\Xi_{cc}$-baryon masses calculated in the
NRQCD sum rules for the formfactors $F_1$ and $F_2$ in the scheme of moments
for the spectral densities.}
\label{mcc}
\end{figure}

\begin{figure}[ph]
\begin{center}
\begin{picture}(100,80)
\put(0,0){\epsfxsize=10cm \epsfbox{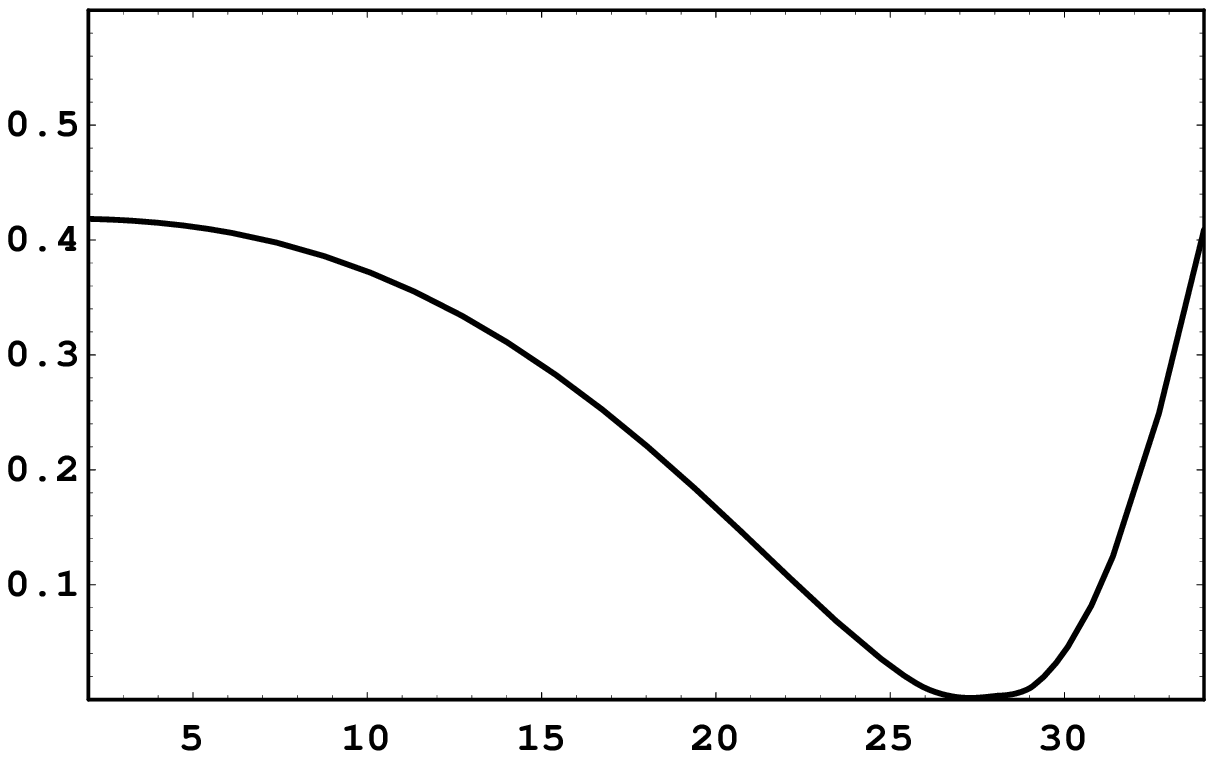}}
\put(100,0){$n$}
\put(0,70){$\Delta M_{\Xi_{bc}}$, GeV}
\end{picture}
\end{center}
\caption{The difference between the $\Xi_{bc}$-baryon masses calculated in the
NRQCD sum rules for the formfactors $F_1$ and $F_2$ in the scheme of moments
for the spectral densities.}
\label{mbc}
\end{figure}

\begin{figure}[ph]
\begin{center}
\begin{picture}(100,80)
\put(0,0){\epsfxsize=10cm \epsfbox{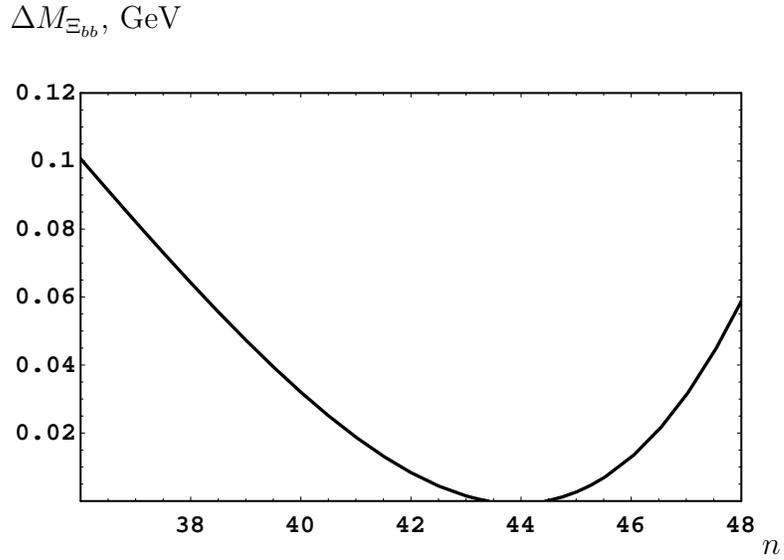}}
\put(100,0){$n$}
\put(0,70){$\Delta M_{\Xi_{bb}}$, GeV}
\end{picture}
\end{center}
\caption{The difference between the $\Xi_{bb}$-baryon masses calculated in the
NRQCD sum rules for the formfactors $F_1$ and $F_2$ in the scheme of moments
for the spectral densities.}
\label{mbb}
\end{figure}

\begin{figure}[th]
\begin{center}
\begin{picture}(100,80)
\put(0,0){\epsfxsize=10cm \epsfbox{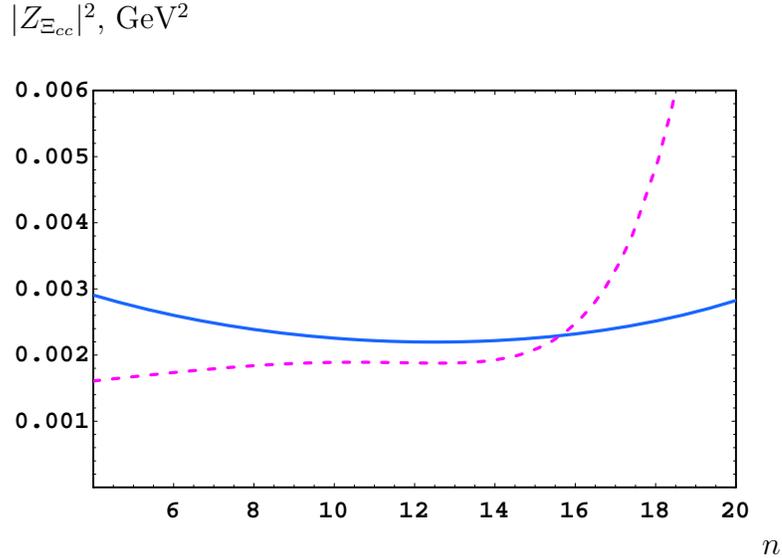}}
\put(100,0){$n$}
\put(0,70){$|Z_{\Xi_{cc}}|^2$, GeV$^2$}
\end{picture}
\end{center}
\caption{The couplings $|Z_{\Xi_{cc}}^{(1,2)}|^2$ of $\Xi_{cc}$-baryon
calculated in the NRQCD sum rules for the formfactors $F_1$ and $F_2$ (solid
and dashed lines, correspondingly) in the scheme of moments for the spectral
densities.}
\label{zcc}
\end{figure}

\begin{figure}[ph]
\begin{center}
\begin{picture}(100,80)
\put(0,0){\epsfxsize=10cm \epsfbox{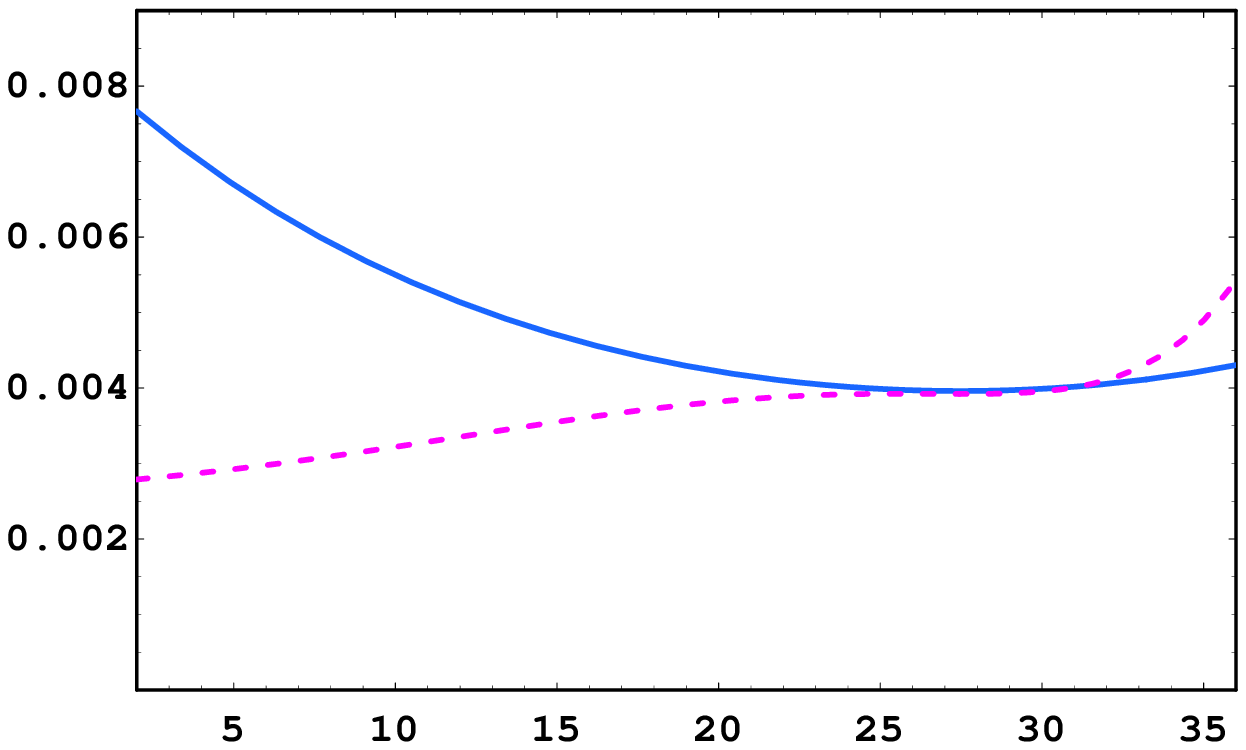}}
\put(100,0){$n$}
\put(0,70){$|Z_{\Xi_{bc}}|^2$, GeV$^2$}
\end{picture}
\end{center}
\caption{The couplings $|Z_{\Xi_{bc}}^{(1,2)}|^2$ of $\Xi_{bc}$-baryon
calculated in the NRQCD sum rules for the formfactors $F_1$ and $F_2$ (solid
and dashed lines, correspondingly) in the scheme of moments for the spectral
densities.}
\label{zbc}
\end{figure}

\begin{figure}[ph]
\begin{center}
\begin{picture}(100,80)
\put(0,0){\epsfxsize=10cm \epsfbox{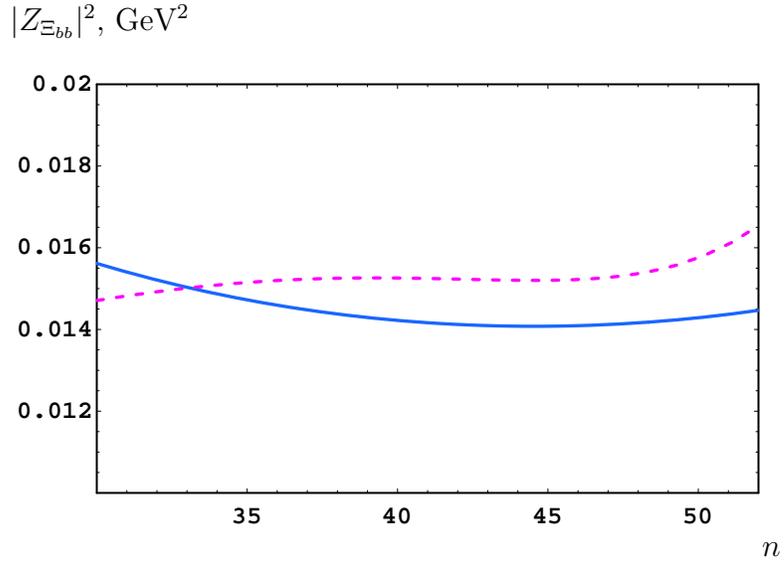}}
\put(100,0){$n$}
\put(0,70){$|Z_{\Xi_{bb}}|^2$, GeV$^2$}
\end{picture}
\end{center}
\caption{The couplings $|Z_{\Xi_{bb}}^{(1,2)}|^2$ of $\Xi_{bb}$-baryon
calculated in the NRQCD sum rules for the formfactors $F_1$ and $F_2$ (solid
and dashed lines, correspondingly) in the scheme of moments for the spectral
densities.}
\label{zbb}
\end{figure}

Figs. \ref{zcc}-\ref{zbb} show the dependence of baryon couplings calculated in
the moment scheme of NRQCD sum rules. We find that the stability regions for
$|Z_\Xi|^2$ determined from the $F_1$ and $F_2$ correlators coincide with those
of the mass differences. So, the baryon couplings in NRQCD are equal to
\begin{eqnarray}
|Z^{\rm NR}_{\Xi_{cc}}|^2 &=& (1.7\pm 0.3)\cdot 10^{-3}\; {\rm GeV}^6,
\nonumber\\
|Z^{\rm NR}_{\Xi_{bc}}|^2 &=& (3.7\pm 0.5)\cdot 10^{-3}\; {\rm GeV}^6,
\label{srstat}\\
|Z^{\rm NR}_{\Xi_{bb}}|^2 &=& (1.5\pm 0.3)\cdot 10^{-2}\; {\rm GeV}^6.\nonumber
\end{eqnarray}
The values given above have to be multiplied by the Wilson coefficients coming
from the expansion of full QCD operators in terms of NRQCD fields, as they have
been estimated by use of corresponding anomalous dimensions. This procedure
results in
\begin{eqnarray}
|Z_{\Xi_{cc}}|^2 &=& (6.5\pm 1.2)\cdot 10^{-3}\; {\rm GeV}^6,\nonumber\\
|Z_{\Xi_{bc}}|^2 &=& (8.5\pm 0.9)\cdot 10^{-3}\; {\rm GeV}^6,\\
|Z_{\Xi_{bb}}|^2 &=& (2.5\pm 0.3)\cdot 10^{-2}\; {\rm GeV}^6.\nonumber
\end{eqnarray}
which are inside the regions given in the analysis of sum rules in full QCD.

For the sake of comparison, we derive the relation between the baryon coupling
and the wave function of doubly heavy baryon evaluated in the framework of
potential model in the approximation of quark-diquark factorization. So, we
find
\begin{equation}
|Z^{\rm PM}| = 2 \sqrt{3} |\Psi_d(0)\cdot \Psi_l(0)|,
\end{equation}
where $\Psi_d(0)$ and $\Psi_l(0)$ denote the wave functions at the origin for
the doubly heavy diquark and light quark-diquark systems, respectively. In the
approximation used, the values of $\Psi(0)$ were calculated in \cite{pot} in
the potential by Buchm\" uller--Tye \cite{BT}, so that
\begin{eqnarray}
\sqrt{4\pi}\;|\Psi_l(0)|    &=& 0.53\; {\rm GeV}^{3/2},\nonumber\\
\sqrt{4\pi}|\Psi_{cc}(0)| &=& 0.53\; {\rm GeV}^{3/2},\nonumber\\
\sqrt{4\pi}|\Psi_{bc}(0)| &=& 0.73\; {\rm GeV}^{3/2},\nonumber\\
\sqrt{4\pi}|\Psi_{bb}(0)| &=& 1.35\; {\rm GeV}^{3/2}.\nonumber
\end{eqnarray}
These parameters result in the estimates in the static limit of potential
models
\begin{eqnarray}
|Z^{\rm PM}_{\Xi_{cc}}|^2 &=& 6.0\cdot 10^{-3}\; {\rm GeV}^6,\nonumber\\
|Z^{\rm PM}_{\Xi_{bc}}|^2 &=& 1.1\cdot 10^{-2}\; {\rm GeV}^6,\\
|Z^{\rm PM}_{\Xi_{bb}}|^2 &=& 3.9\cdot 10^{-2}\; {\rm GeV}^6.\nonumber
\end{eqnarray}
As it is well known from the analysis of heavy-light mesons, the approximation
of potential models for the static limit of coupling results in the
overestimation of leptonic constants of heavy mesons with the single heavy
quark because of the large corrections in the expansion of corresponding
currents through the effective static fields. So, the contribution by the
higher dimensional operators is essential, and it leads to the suppression
factor of about 1/2 for the couplings. The introduction of this factor allows
us to calculate the leptonic constants of heavy mesons with the same wave
function of light quark
$$
f_D=185\; {\rm MeV},\;\;\;
f_B=115\; {\rm MeV},\;\;\;
$$
which are quite reasonable estimates, being in agreement with the results of
QCD sum rules \cite{nar}. Then, we use the same factor for the renormalization
of baryon couplings and find
\begin{eqnarray}
|\bar Z^{\rm PM}_{\Xi_{cc}}|^2 &=& 1.5\cdot 10^{-3}\; {\rm GeV}^6,\nonumber\\
|\bar Z^{\rm PM}_{\Xi_{bc}}|^2 &=& 2.8\cdot 10^{-3}\; {\rm GeV}^6,
\label{potstat}\\
|\bar Z^{\rm PM}_{\Xi_{bb}}|^2 &=& 9.8\cdot 10^{-3}\; {\rm GeV}^6.\nonumber
\end{eqnarray}
Comparing the values in (\ref{potstat}) with ones in (\ref{srstat}), we see a
good agreement of NRQCD results on the baryon couplings with estimates
of potential models improved by the correction factor, if we take into account
the ordinary accuracy about 30\% for the phenomenological quark models.

Finally, we suppose that the corrections coming from the higher orders of NRQCD
expansion to the baryon couplings are not greater than 15\%, since the diquark
masses are quite large.

Thus, we obtain the reliable estimates of masses and couplings for the doubly
heavy baryons in the framework of NRQCD sum rules.

\section{Conclusion}

We have considered the NRQCD sum rules for the two-point correlators of
baryonic currents with two heavy quarks. The nonrelativistic approximation for
the heavy quark fields allows one to fix the structure of currents and to take
into account the coulomb-like interactions inside the doubly heavy diquark.
Moreover, we introduce the higher order operators responsible for the
quark-gluon condensates to get the convergency of sum rule method for two
scalar functions of correlators. To the leading approximation, including the
perturbative term and the contributions of quark and gluon condensates, the
correlators of three-quarks state and the doubly heavy diquark are factorized
in separate functions, so that the sum rules result in the different values of
masses and couplings. This fact indicates the divergency of approach unless the
product of quark and gluon condensates and the mixed condensate are taken into
account. Then, the interaction of two heavy quarks and light quark destroy the
factorization, which allows one to get meaningful estimates of masses and
couplings. Moreover, we have also calculated the binding energy of doubly heavy
diquark, which is in a good agreement with the estimates in the framework of
potential models.

Thus, the NRQCD sum rules allow us to improve the analysis of masses and
couplings for the doubly heavy baryons and to obtain reliable results.

This work is in part supported by the Russian Foundation for Basic Research,
grants 99-02-16558 and 96-15-96575. The work of A.I.Onishchenko was supported,
in part, by International Center of Fundamental Physics in Moscow,
International
Science Foundation, and INTAS-RFBR-95I1300 grants.

\end{document}